\documentclass[12pt]{iopart}
\usepackage{iopams}  
\usepackage{graphicx}
\usepackage{dcolumn}
\usepackage{bm}
\usepackage[dvips]{color}
\usepackage{xspace}
\usepackage{theorem} 
\usepackage{latexsym} 
\usepackage[mathscr]{eucal} 
\usepackage{color}

\newcommand{\lv}{\left \vert}
\newcommand{\rv}{\right \vert}
\newcommand{\la}{\left \langle}
\newcommand{\ra}{\right \rangle}
\newcommand{\ket}[1]{\lv #1 \ra}
\newcommand{\bra}[1]{\la #1 \rv}

\newcommand{\ie}{\textit{i.e.}\xspace}

\newcommand{\A}{\mathrm{A}}
\newcommand{\B}{\mathrm{B}}

\begin{document}

\title{Delocalization power of global unitary operations on quantum information}

\author{A Soeda$^{1}$ and M Murao$^{1,2}$}

\address{
  $^1$Department of Physics, Graduate School of Science, The University of Tokyo, Tokyo 113-0033, Japan}
\address{
  $^2$Institute for Nano Quantum Information Electronics, The University of Tokyo, Tokyo 113-0033, Japan }
\ead{soeda@eve.phys.s.u-tokyo.ac.jp}

\begin{abstract}
We investigate how originally localized two pieces of quantum information represented by a tensor product of two unknown qudit states are delocalized by performing two-qudit global unitary operations.  
To characterize the delocalization power of global unitary operations on quantum information, we analyze the necessary and sufficient condition to deterministically relocalize one of the two pieces of quantum information to its original Hilbert space by using only LOCC.   We prove that this LOCC one-piece relocalization is possible if and only if the global unitary operation is local unitary equivalent to a controlled-unitary operation.  The delocalization power and the entangling power characterize different non-local properties of global unitary operations.
\end{abstract}

\submitto{\NJP}
\maketitle

\section{Introduction}

Quantum information processing (QIP) aims to achieve more efficient information processing or to perform tasks that are not classically possible, by utilizing the superposition principle of quantum states.
Extending our knowledge of what QIP can provide is a driving force of developments in quantum technologies and it also reveals the fundamental difference between quantum and classical systems.    Understanding what makes QIP outperform its classical counterpart is essential for understanding yet unknown potentials of QIP.

Information processing can be interpreted as operations mapping from input information into output information in both classical and quantum cases.  In both cases,  operations over several bits or qubits are essential for information processing.  In QIP, operations over several qubits are called {\it global} (or non-local) operations.  Global operations can generate entanglement, one of the most distinguishable characters of quantum mechanics from classical mechanics, whereas local operations over single qubits cannot generate entanglement even adding classical correlations.   Entanglement has been investigated in the local operation and classical communication (LOCC) paradigm \cite{LOCC}.

Although the entanglement of quantum states provides many clues for understanding the superiority of QIP \cite{E91,Shor,metrology}, it has also turned out that entanglement alone is not sufficient for understanding every aspect of QIP.  The Gottesman-Knill theorem \cite{GK} implies that classical computers can efficiently simulate a certain type of QIP generating entanglement.  On the other hand, in a model of quantum computation called the deterministic quantum computation with one pure quantum bit (DQC1) \cite{DQC1}, there are some cases that quantum speedup can be achieved without generating entanglement.  We need another clue for understanding advantages of QIP.

One of the essential differences in QIP and classical counterpart is that QIP sometimes involves global operations on {\it unknown} input states, namely, arbitrary superpositions of quantum states where their superposition coefficients are unknown.  Quantum teleportation \cite{Teleportation} and quantum error corrections \cite{EC} are typical examples.  We call such unknown states as {\it quantum information} in this paper.  Quantum information cannot be measured (i.e. estimating the unknown coefficients by finite measurements) perfectly and cannot be copied perfectly either \cite{No-cloning}.  In contrast, the classical information in QIP can be encoded in a set of {\it known} orthogonal states and it can be perfectly measured.  Classical information can be also obtained by the result of measurements in QIP.  QIP can be analyzed by investigating how input quantum information is transformed to output quantum information due to the global operations.   Therefore, the evaluation of the effects of the global operation on quantum information is desired.

For evaluation, the simplest global operations in QIP are global unitary operations.   One way of evaluating global unitary operations is by their entangling power \cite{KrausCirac}.  The entangling power is the maximum amount of entanglement generated by a unitary operation among all possible quantum input states.  It gives a way to quantify global unitary operations and is useful for finding the lower bound of the entanglement resource required to perform the unitary operation between distributed parties.  But it evaluates the maximum value achieved for {\it certain} states, and its operational meaning in terms of quantum information is not clear.

In this paper, we present an approach for this problem by introducing the concept of delocalization
of initially localized two ``pieces'' of quantum information by a global unitary operation $U$ and investigating the power of delocalization.  To characterize how ``powerful'' is the delocalization effect on two pieces of quantum information caused by the global unitary operation, we introduce another concept called relocalization.   Relocalization is a process to {\it deterministically} transform {\it at least} one of the pieces of delocalized quantum information into its minimal Hilbert space.
Our idea is to evaluate the power of delocalization of global unitary operations by analyzing non-local properties of the quantum operations required for relocalizing one of the two pieces of delocalized quantum information.

We define two classes of global unitary operations on the two-qudit (quantum $d$-level) Hilbert space in terms of the delocalization power characterized by one-piece relocalization.   For the first class of global unitary operations, one piece of quantum information can be relocalized by only using LOCC. For the second class, one-piece relocalization cannot be performed by LOCC, namely, it requires global operations.  In this paper, we show that the first class consists of controlled-unitary operations and their local unitary equivalent operations.   Any other global unitary operations fall into the second class.   Since the second class requires more non-local resources than the first class, our classification also gives a way to rank the delocalization power of global unitary operations.

We note that a related problem has been investigated by Gregoratti and Werner \cite{GW} (quantum lost and found).  Their problem can be also interpreted to find a condition for delocalized quantum information to be relocalized by just using LOCC.  In their setting, one of the two qudits is chosen as the environment and the initial state of the environment is fixed to be a known pure state.   It corresponds to a special situation of our case where only one piece of quantum information is delocalized in the two-qudit Hilbert space.  In their problem, measurements in LOCC are only allowed on the environmental qudit.  This restriction on the measurement is equivalent to restricting general LOCC to more limited {\it one-way} LOCC.   They have investigated the condition on the completely positive and trace-preserving (CPTP) map of the non-environmental qudit alone, not global unitary operations, for the delocalized piece of quantum information to be relocalizable by LOCC.  They have shown that the necessary and sufficient condition of the CPTP map is that the map is a random unitary channel.   Ogata and Murao \cite{OM} have also considered a related problem to find a necessary and sufficient condition for a restricted class of delocalized two pieces of {\it qubit} quantum information in the two-qubit Hilbert space to be relocalized by LOCC.  In their work, they considered general LOCC, but their LOCC operations required partial knowledge of one of the qubit states.   Thus, the main difference between these previous works and our work is that we treat full two pieces of input quantum information to evaluate the delocalization power of global unitary operations. 

This paper is organized by the followings.  In Section 2, we define the delocalization and relocalization of quantum information and introduce the classification of the two-qudit global unitary operations in terms of the delocalization power evaluated by LOCC one-piece relocalizability.  In Section 3, we show that if a general LOCC one-piece relocalization protocol exists, then it can be reduced to one-way LOCC.  In  Section 4, we show that if a one-way LOCC one-piece relocalization protocol using a general measurement exists, then there also exists a one-way LOCC protocol using a projective measurement. And we derive the necessary and sufficient condition for two-qudit global unitary operations to be LOCC one-piece relocalizable.    In Section 5, we give conclusion and discussions. 

\section{Delocalization and relocalization}
\begin{figure}[t]
\begin{center}
\includegraphics[width=10cm]{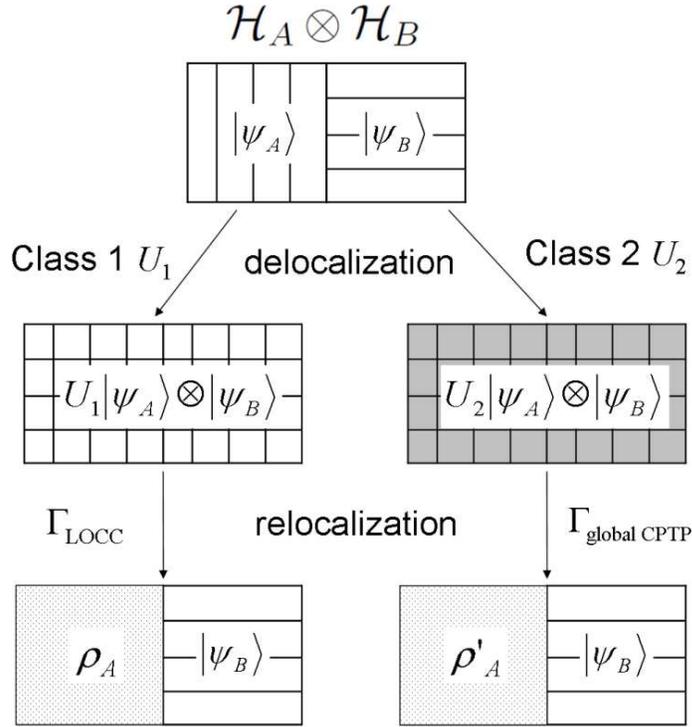}
\caption{Schematic picture of the concept of delocalization and LOCC one-piece relocalization.  The rectangle on the top consists of two squares, representing the two-qudit Hilbert space $\mathcal{H}_\A \otimes \mathcal{H}_\B$.  The left square with vertical lines represents Alice's Hilbert space $\mathcal{H}_\A$, in which a pure state $\ket{\psi_\A}$ is localized.  Similarly, the right square with the vertical lines represents Bob's Hilbert space $\mathcal{H}_\B$ with $\ket{\psi_\B}$ being localized.  The arrows on the left hand side indicate the delocalization of the two pieces of quantum information by a Class 1 global unitary operation $U_1$ followed by the one-piece relocalization of Bob's quantum information using a LOCC map $\Gamma_{\mathrm{LOCC}}$.  The arrows on the right hand side correspond to the delocalization by a Class 2 global unitary operaiton $U_2$, which is relocalized by a \textit{global} CPTP map $\Gamma_{\mathrm{global~CPTP}}$.  An appropriate entangled resource must be supplied on top of LOCC to implement $\Gamma_{\mathrm{global~CPTP}}$.  In both cases, the two pieces of localized quantum information spread across the whole two-qudit Hilbert space (becoming ``intertwined'' with each other) and Alice's piece of quantum information is sacrificed.}
\end{center}
\end{figure}

Before introducing delocalization and relocalization, we present our notion of ``pieces'' of quantum information.  We consider an unknown state of a qudit described by $\ket{\psi}=\sum_{i=0}^{d-1} \alpha_i \ket{i}$ where $\{ \ket{i} \}$ is a basis of the $d$-dimensional
Hilbert space $\mathcal{H}=\mathbb{C}^d$ and $\{ \alpha_i \}$ is a set of arbitrary coefficients normalized by $\sum{| \alpha_i |^2}=1$.   We interpret this unknown state as a situation that a {\it piece} of quantum information of a qudit is {\it localized} in its minimal Hilbert space $\mathcal{H}$.

We consider a situation that each one of  two parties named Alice and Bob possesses a piece of quantum information. Let $\mathcal{H}_\A$ and $\mathcal{H}_\B$ denote the Hilbert space describing Alice's and Bob's qudit, respectively.   The joint state describing the two pieces of quantum information is given by a tensor product of two qudit states $\ket{\psi_{\mathrm{AB}}}=\ket{\psi_\A} \otimes \ket{\psi_\B}$, where $\ket{\psi_\A} \in \mathcal{H}_\A $ and $\ket{\psi_\B} \in \mathcal{H}_\B$.   Note that, by definition of calling the state to be quantum information, Alice and Bob do not know the state, although we assume they know that the state $\ket{\psi_{\mathrm{AB}}}$ consists of a tensor product of two pure states.

Now we introduce the delocalization of quantum information.   In this paper, we use the word global unitary operations to denote unitary operations acting globally on the two-qudit Hilbert space $\mathcal{H}_{\mathrm{AB}}=\mathcal{H}_\A \otimes \mathcal{H}_\B$, not including the tensor products of single-qudit unitary operations.  Performing a two-qudit global unitary operation $U$ on a tensor product state  $\ket{\psi_{\mathrm{AB}}}$ generally gives an entangled state $U\ket{\psi_\A}\otimes\ket{\psi_\B}$ in the joint Hilbert space $\mathcal{H}_{\mathrm{AB}}$.  We regard that the both pieces of quantum information are  {\it delocalized} over the same joint Hilbert space due to the global unitary operation $U$.

Relocalization is a process to deterministically transform at least one of the pieces of delocalized quantum information over  $\mathcal{H}_{\mathrm{AB}}$ into its minimal Hilbert space  $\mathcal{H}_\A$ for Alice's piece of quantum information $\ket{\psi_\A}$ or $\mathcal{H}_\B$ for Bob's piece of quantum information $\ket{\psi_\B}$.   To relocalize two pieces of delocalized quantum information perfectly to their original form, we need to perform the conjugate unitary operation $U^\dagger$.    However, if we only require {\it one piece} of delocalized quantum information to be relocalized perfectly, this one-piece relocalization can be done by more general quantum operations described by completely positive trace preserving (CPTP) maps, not necessarily the global unitary operation $U^\dagger$.

In the rest of the paper, we analyze the case when Bob's piece of quantum information $\ket{\psi_\B}$ is to be relocalized to $\mathcal{H}_\B$.   Denoting the CPTP map of a deterministic one-piece relocalization for a global unitary operation $U$ to be $\Gamma_U$, its action is given by
\begin{eqnarray}
\Gamma_U \left ( U \ket{\psi_\A} \bra{\psi_\A} \otimes \ket{\psi_\B} \bra{\psi_\B}  U^\dagger \right )= \rho_\A  \otimes \ket{\psi_\B} \bra{\psi_\B}
\end{eqnarray}
for arbitrary $\ket{\psi_\A}$ and $\ket{\psi_\B}$, where $\rho_\A \in \mathcal{S}(\mathcal{H}_\A)$ is a density matrix independent of $\ket{\psi_\B}$.

We analyze how strong is the delocalization power of the global unitary operation $U$ on quantum information, by investigating how non-local the one-piece relocalization CPTP map $\Gamma_U$ should be.   Particularly, in this paper, we classify the global unitary operations $U$ by investigating the necessary and sufficient condition for a one-piece relocalization CPTP map $\Gamma_U$ to be achieved by LOCC.  All global unitary operations can be divided into the following two classes:
\begin{description}
 \item[Class 1:]  Global unitary operations that are one-piece relocalizable by LOCC.
 \item[Class 2:]  Global unitary operations that are {\it not} one-piece relocalizable by LOCC
\end{description}
These classes give a ranking of global unitary operations by their delocalization power.

Showing that a given global unitary operation belongs to Class 1 is relatively easy.  We just need to show a construction of a LOCC protocol for one-piece relocalization.   On the other hand, it is more difficult to prove that a given global unitary operation belongs to Class 2, because it must be shown that one-piece relocalization is impossible by any LOCC protocol.

We prove this classification problem by the following manner.   First, we derive a necessary condition on the LOCC protocol for one-piece relocalization for a particular input state by introducing accumulated operators representing LOCC.  We show that this condition restricts the operators representing the protocol and this restriction implies that any LOCC one-piece relocalization protocol can be reduced to a {\it one-way} LOCC protocol.   Such one-way LOCC protocols are shown to be decomposed into Alice's general measurement and Bob's local unitary operation conditional to the Alice's measurement outcome.  By reducing the one-way LOCC protocol with a general measurement into the one with a projective measurement, we prove that a global unitary operation is one-piece relocalizable if and only if it is a controlled-unitary operation and its local unitary equivalent, where the control qudit is Alice's and the target Bob's.

\section{Reduction to one-way LOCC}

We denote the LOCC one-piece relocalization CPTP map for a global unitary operation $U$ to be ${\Gamma}_U^L$. Suppose that each of Alice and Bob has an extra ancilla qudit.   Although the relocalization CPTP map is only required to relocalize one piece of quantum information, the linearity of CPTP maps guarantees that such relocalization map also recovers the state of one of the two input qudits even when each of the qudits is entangled to the ancilla qudit.

Assume that Alice's input qudit is maximally entangled to an ancilla qudit and Bob's input qudit is also maximally entangled to another ancilla qudit.  The reduced density matrix of the joint state of Alice and Bob's input qudits is given by the tensor product of two completely mixed states, which is 
described by ${\mathbb I}/d \otimes {\mathbb I}/d$, where ${\mathbb I}$ denotes the identity operator on $d$-dimensional Hilbert space.  Then the action of $\Gamma_U^{L}$ should be given by
\begin{equation}
\Gamma_U^{L}  (U (\frac{{\mathbb I}}{d} \otimes \frac{{\mathbb I}}{d}) U^\dagger) = \rho_\A \otimes \frac{{\mathbb I}}{d} \label{xer}
\end{equation}
for some density matrix $\rho_\A$.  The left hand side of (\ref{xer}) can be simplified because the tensor product of the identity operators commutes with the global unitary operator $U$.  Since the CPTP map is independent of $U$ for this special initial state, we denote $\Gamma^{L} \equiv \Gamma_{U={\mathbb I}}^{L}$ and simplify the equation to
\begin{equation}
\Gamma^{L} (\frac{{\mathbb I}}{d} \otimes \frac{{\mathbb I}}{d}) = \rho_\A \otimes \frac{1}{d}  {\mathbb I}.
\label{IdentityLOCCrelocalizationMap}
\end{equation}

In general, LOCC protocols consist of a series of local measurements on the local Hilbert space conditional to the classical information of the measurement outcomes of the previous local measurements, except for the first turn and the last turn of LOCC, where the first local measurement is not conditional to the previous measurements by definition, and the last turn is not necessary to be finished by a measurement, but a unitary operation.

We describe a LOCC protocol by using a set of ``accumulated'' operators consisting of a tensor product of Alice's accumulated operations $M^{\vec{n}_i}$ acting on $\mathcal{H}_\A$ and Bob's accumulated operations $K^{\vec{n}_i}$ acting on $\mathcal{H}_\B$ where the superscript $\vec{n}_i$ indicates the dependence on all the measurement outcomes up to the $i$-th turn.   The measurement outcome of the $k$-th ($1 \leq k \leq i$) turn is represented by the $k$-th  component $n_k$ of the vector $\vec{n}_i= (n_1, n_2, \ldots, n_i)$.   The set of the vectors $\{ \vec{n}_i \}$ denotes all possible sequences of measurement outcomes Alice and Bob have obtained up to the $i$-th turn. Note that $\vec{n}_i$ indicates a ``branch'' of measurement outcomes, and does not only indicate the measurement outcome of the $i$-th turn.  The vector $\vec{n}_i$ and its element $n_i$ should not be confused.   

By denoting the conditional measurement operator for the $i$-th measurement performed by Bob for a measurement branch $\vec{n}_i$ to be $\mathcal{K}^{\vec{n}_i | \vec{n}_{i-1}}$, the relationship between two accumulated operators $K^{\vec{n}_i}$ and $K^{\vec{n}_{i-1}}$ is given by 
\begin{eqnarray}
K^{\vec{n}_i} = \mathcal{K}^{\vec{n}_i | \vec{n}_{i-1}} K^{\vec{n}_{i-1}}.
\label{acrelation}
\end{eqnarray}
A similar relation holds for $M^{\vec{n}_i} $ if the $i$-th measurement is performed by Alice.  By iterating this procedure, the accumulated operators can be decomposed to products of conditional local measurements alternatingly performed by Alice and Bob, and this gives a standard notation of LOCC \cite{Horodecki-LOCC definition}.   

Using the accumulated operators, a general $N$-turn LOCC maps a density matrix $\rho_{\mathrm{AB}} \in\mathcal{S} (\mathcal{H}_\A \otimes \mathcal{H}_\B)$ to
\begin{equation}
\rho_{\mathrm{AB}}
\rightarrow
\sum_{\vec{n}_N} (M^{\vec{n}_N}\otimes K^{\vec{n}_N}) \rho_{\mathrm{AB}} (M^{\vec{n}_N}\otimes K^{\vec{n}_N})^\dag.
\end{equation}
To guarantee the trace preserving property, the completeness relation of the accumulated operators $M^{\vec{n}_k}$ and $K^{\vec{n}_k}$, namely,
\begin{equation}
\sum_{n_i} M^{\vec{n}_i \dag} M^{\vec{n}_i} =  M^{\vec{n}_{i-1} \dag} M^{\vec{n}_{i-1}}
\end{equation}
and 
\begin{equation}
\sum_{n_i} K^{\vec{n}_i \dag}K^{\vec{n}_i} =  K^{\vec{n}_{i-1} \dag} K^{\vec{n}_{i-1}},
\label{compK}
\end{equation} 
should be satisfied for all $1 \leq i \leq N$.

Back to LOCC one-piece relocalization, its $N$-turn LOCC CPTP map $\Gamma^L$ given by (\ref{IdentityLOCCrelocalizationMap}) can be represented by using the accumulated operators by
\begin{eqnarray}
\Gamma^{L} (\frac{{\mathbb I}}{d} \otimes \frac{{\mathbb I}}{d}) &=&
\sum_{\vec{n}_N}  (M^{\vec{n}_N}\otimes K^{\vec{n}_N}) \frac{{\mathbb I}}{d} \otimes \frac{{\mathbb I}}{d}  (M^{\vec{n}_N} \otimes K^{\vec{n}_N})^\dagger \nonumber \\
&=&
\frac{1}{d^2} \sum_{\vec{n}_N}  M^{\vec{n}_N} M^{\vec{n}_N^\dagger} \otimes K^{\vec{n}_N} K^{\vec{n}_N^\dagger} \nonumber \\
&=& \rho_A \otimes \frac{1}{d}  {\mathbb I}.
\end{eqnarray}
Since Bob's qudit should be maximally entangled to his ancilla after the one-piece relocalization, 
\begin{equation}
K^{\vec{n}_N} K^{\vec{n}_N \dag} \propto {\mathbb I} \label{proptoI}
\end{equation}
should be satisfied for {\it all} possible branches of outcomes $\vec{n}_N$.  Therefore, Bob's accumulated operator has to be proportional to a unitary operation and can be written as $K^{\vec{n}_N } =\sqrt{p^{\vec{n}_N}} u_\B^{\vec{n}_N}$ for all $\vec{n}_N$, where $p^{\vec{n}_N}$ is a real number satisfying $0 \leq p^{\vec{n}_N} \leq 1, \sum_{\vec{n}_N} {p^{\vec{n}_N}}=1$ and $u_\B^{\vec{n}_N}$ is a local unitary operator.   This Bob's accumulated operator can be interpreted to perform a random unitary operation $u_\B^{\vec{n}_N}$ depending on the branch ${\vec{n}_N}$ with probability $p^{\vec{n}_N}$.

By using (\ref{compK}), we can show that $K^{\vec{n}_{N-1} \dag} K^{\vec{n}_{N-1}}  \propto \mathbb{I}$, and by iterating this procedure we obtain $K^{\vec{n}_i} \propto u_\B^{\vec{n}_i}$ for all $1 \leq i \leq N$.  Then Bob's conditional measurement operator $\mathcal{K}^{\vec{n}_i | \vec{n}_{i-1}} $ for all $\vec{n}_i$ and $i$ should also be proportional to a unitary operator.   Thus, all Bob's operations in LOCC one-piece relocalization have to be random unitary operations that depend on the sequence of the measurement outcomes up to the previous turn.

When Bob's operation in a LOCC protocol at a particular turn is a random unitary operation, Alice's measurements in the subsequent turns do not depend on this Bob's outcome.  In this case, Bob does not need to send any classical information to Alice.  Therefore, any LOCC one-piece relocalization protocol does not require the communication from Bob to Alice, and can be reduced to a one-way LOCC protocol. 

\section{Relocalization by one-way LOCC}

By redefining the combination of multiple measurements on Alice's qudit as a single measurement, we can reduce our problem to a one-way LOCC protocol using only one measurement.
Then the accumulated operator of the one-way LOCC one-piece relocalzation protocol for a global unitary operator $U$ is given by
\begin{equation}
 M_U^n \otimes u_U^n,
\end{equation}
where $M_U^n$ is a measurement operator corresponding to the measurement outcome $n$ and $u_U^n$ is Bob's conditional local unitary operation.   Note that a vector superscript ${\vec{n}}$ reduced to a single variable $n$ denoting Alice' measurement outcome in the first turn.

Using this accumulated operator,  the action of the one-way LOCC CPTP map of (deterministic) one-piece relocalization has to satisfy
\begin{eqnarray}
(M_U^n \otimes u_U^n ) U \ket{\psi_\A} \bra{\psi_\A} \otimes \ket{\psi_\B} \bra{\psi_\B} U^\dagger (M_U^n \otimes u_U^n )^\dagger \nonumber \\
= \rho_{\A,U}^n \otimes \ket{\psi_\B} \bra{\psi_\B}
\end{eqnarray}
for all $n$.   Since $\ket{\psi_\A}$ and $\ket{\psi_\B}$ can be arbitrary, $ (M_U^n \otimes u_U^n) U = M_U^{'n} \otimes \mathbb{I}$ should hold, where $M_U^{'n}$ is an operator satisfying $M_U^{'n} \ket{\psi_\A} \bra{\psi_\A} M_U^{'n \dagger} = \rho_{\A,U}^n$.   Then we have $M_U^{n} M_U^{n \dagger} =  M_U^{'n} M_U^{'n \dagger}$, which implies that $M_U^{'n} = M_U^{n} v_U^n$ with an appropriate local unitary operator $v_U^n$ of Alice.  Therefore, we obtain a condition for the LOCC one-piece relocalization given by
\begin{equation}
 (M_U^{n}  \otimes u_U^n) U = (M_U^{n}  v_U^n)  \otimes \mathbb{I}.
\label{ppm}
\end{equation}

In general, the measurement operators $M_U^n$ in a one-way LOCC are not necessary to be Hermitian.  However for our analysis of LOCC one-piece relocalization, we can always take Hermitian measurement operators without loss of generality.  To see this, we denote the polar decomposition of $M_U^n$ by
\begin{equation}
M_U^{n} = w_U^n |M_U^{n}|,
\label{polar_d} 
\end{equation}
where $|M_U^{n}| = \sqrt{M_U^{n \dagger} M_U^{n} }$ and $w_U^n$ is a unitary operator.  Then any non-Hermitian measurement operator $M_U^n$ can be represented by an Hermitian measurement operator $|M_U^{n}|$ followed by an appropriate local unitary operation $w_U^n$.   The additional local unitary operation $w_U^n$ does not affect the classification in terms of LOCC one-piece relocalization.

The Hermitian measurement operators $M_U^n$ can be written by
\begin{equation}
 M_U^n = \sum_k m_U^{n,k} P_U^{n,k} \label{spec_d}
\end{equation}
where $m_U^{n,k}$ represents an eigenvalue of $M_U^n$ and $P_U^{n,k}$ is a rank-1 projector of whose range is the eigenspace corresponding to $m_U^{n,k}$.  By multiplying $P_U^{n,k} \otimes u_U^{n \dagger}$ from left on (\ref{ppm}) and using (\ref{polar_d}) and (\ref{spec_d}),
we obtain
\begin{equation}
 (P_U^{n,k} \otimes \mathbb{I} ) U = (P_U^{n,k} v_U^n) \otimes u_U^{n \dagger}.
\end{equation}
Taking the summation over $k$ and introducing a projector given by $P_U^n = \sum_k P_U^{n,k}$ of whose range is the support of $M_U^n$, we obtain a simplified condition of LOCC one-piece relocalization given by
\begin{equation}
 (P_U^n \otimes \mathbb{I}) U = (P_U^n v_U^n) \otimes u_U^{n \dagger}. \label{qqc}
\end{equation}
By multiplying its complex conjugate of (\ref{qqc}) from right, we have
\begin{equation}
 P_U^n P_U^m \otimes \mathbb{I} = (P_U^n v_U^n v_U^{m \dag} P_U^m) \otimes (u_U^{n \dag} u_U^m).
\label{pnm}
\end{equation}

First, we consider a special case where Alice's measurement is given by a projective measurement satisfying $M_U^n=P_U^n$ and $P_U^n P_U^m = \delta_{n,m} P_U^n$.   In this case, from (\ref{pnm}), the condition for  the operators acting on Alice's Hilbert space is given by $P_U^n v_U^n v_U^{m \dag} P_U^m = \delta_{n,m} \mathbb{I}$.  This condition implies that there exists a unitary operator $v_U$ satisfying $P_U^n v_U = P_U^n   v_U^n$ and $P_U^m  v_U = P_U^m  v_U^m$.  On the other hand, the unitary operators acting on Bob's Hilbert space $u_U^{n \dag} u_U^m$ do not have any restriction.   
Thus we rewrite the condition (\ref{qqc}) into 
\begin{equation}
 (P_U^n \otimes \mathbb{I}) U =  (P_U^n v_U) \otimes u_U^{n \dagger}, \label{projm}
\end{equation}
where Alice's unitary operator $v_U$ does not dependent on $n$ any more. 

By taking summation over $n$ of (\ref{projm}), and using the completeness relation of the projectors $\sum_n P_n = \mathbb{I}$, we obtain the condition for a global unitary operation
\begin{equation}
 U = \left( \sum_n P_U^n  \otimes  u_U^{n \dag}  \right ) \cdot v_U \otimes \mathbb{I}. \label{jik}
\end{equation}
The first part of the right hand side, namely, 
\begin{equation}
\sum_n P_U^n  \otimes u_U^{n \dag}
\label{localunitaryform}
\end{equation}
denotes a controlled-unitary operation where the local unitary operation $u_U^{n \dag}$ is performed on Bob's qudit when the state of Alice's qudit is determined by the projector $P_U^n$.  Therefore, the global unitary operation $U$ should be local unitary equivalent to the controlled-unitary operation given by (\ref{localunitaryform}).   

Next, we consider a more general case where Alice's measurement is a non-projective measurement.  In the following, we are going to show that if $U$ is LOCC one-piece relocalizable by a non-projective measurement, then there always exists a relocalization protocol using a projective measurement.  Thus, we can reduce a one-way LOCC protocol with a general measurement to the one with a projective measurement. 

We consider two projectors $P_U^n$ and $P_U^m$, where $P_U^n P_U^m \neq 0$ for $n \neq m$.   Then the operator $u_U^{n \dag} u_U^m$ acting on Bob's Hilbert space in (\ref{pnm})
has to be equivalent to $\mathbb{I}$ up to a global phase.  We set $u_U^m=\rme^{\rmi \theta_{n,m}} u_U^n$ by choosing an angle $\theta_{n,m}$ where $0 \leq \theta_{n,m} \leq 2 \pi $.   Since $u_U^{n \dag} u_U^m = \rme^{ \rmi \theta_{m,n}} \mathbb{I}$, the condition for the operators acting on Alice's Hilbert space is given by 
\begin{equation}
 P_U^n P_U^m =  P_U^n v_U^n v_U^{m \dag} P_U^m \rme^{\rmi \theta_{m,m}}. 
\label{nmp}
\end{equation}
This equation indicates that there exists a unitary operator $v_U$ such that $P_U^n v_U =P_U^n   v_U^n$ and $P_U^m  v_U = \rme^{\rmi \theta_{n,m}}  P_U^m  v_U^m$.  By using this unitary operator $v_U$ that is independent of $n$,  we can rewrite (\ref{qqc}) by
\begin{equation}
 (P_U^n \otimes \mathbb{I}) U =  (P_U^n v_U) \otimes u_U^{n \dagger} \label{qqcv1}
\end{equation}
and
\begin{equation}
 (P_U^m \otimes \mathbb{I}) U =  (P_U^m v_U) \otimes u_U^{n \dagger}. \label{qqcv2} 
\end{equation}

The ranges of the projectors $P_U^n$ and $P_U^m$ form subspaces.  We denote the subspace specified by the projector $P_U^n$ to be $\mathcal{W}_U^{n}$ and the subspace specified by the projector $P_U^m$ to be $\mathcal{W}_U^{m}$.  We construct the sum-space $\mathcal{W}_U^{n,m}$ of these two subspaces $\mathcal{W}_U^{n}$ and $\mathcal{W}_U^{m}$ and denote the projector of whose range is this sum-space to be $P_U^{n,m}$.  Since the sum-space contains two subspaces $\mathcal{W}_U^{n}$ and $\mathcal{W}_U^{m}$, $P_U^n P_U^{n,m} = P_U^n$ and $P_U^m P_U^{n,m} = P_U^m$ are satisfied. 
From (\ref{qqcv1}) and (\ref{qqcv2}), for any vector $\ket{\phi} \in \mathcal{W}_U^{n,m}$, we have
\begin{equation}
(\bra{\phi} \otimes {\mathbb I}) U =  \bra{\phi} v_U \otimes u_U^{n \dag},
\end{equation}
which proves that
\begin{equation}
 (P_U^{n,m} \otimes {\mathbb I}) U =P_U^{n,m} v_U \otimes u_U^{n \dag}.
\end{equation}
Thus, we can ``combine'' the two non-orthogonal projectors  $P_U^n$ and $P_U^m$  into a single projector  $P_U^{n,m}$.

We repeat this procedure on the set of projectors $\{ P_U^n \}$ until all elements are orthogonal to each other.  By renaming the resulting mutually orthogonal projectors with $P_U^{' k}$, we obtain a set of orthogonal projectors $\{ P_U^{' k} \}$.   The number of elements of $\{ P_U^{' k}  \}$ is strictly smaller than that of $\{ P_U^n \}$ due to the combing procedure.  Thus, we have shown that if $U$ is LOCC one-piece relocalizable by a non-projective measurement, then there always exists a LOCC one-piece relocalization protocol using a projective measurement.  Therefore the necessary condition for a global unitary operation $U$ to be LOCC one-piece relocalizable using general measurements is the same as the case using the projective measurement given by (\ref{jik}), namely, local unitary equivalent to a controlled-unitary operation.   

On the other hand, the sufficient condition for $U$ to be LOCC one-piece relocalizable can be shown by the following construction.   Consider the case where a global unitary operation $U$ is local unitary equivalent to a controlled-unitary operation, i.e. $U = u_\A \otimes u_\B (\sum_m P^m \otimes u^m) v_\A \otimes v_\B$, where $u_\A$, $u_\B$, $v_\A$, $v_\B$ and $u^m$ are local unitary operations and $P^m$ is a projector.   The delocalization of the two pieces of quantum information $\ket{\psi_\A} \otimes \ket{\psi_\B}$ is given by performing the global unitary operation $U$.  Then Alice performs a general measurement described by $\{ P^m u_\A^\dag \}$ on her qudit and communicates the measurement outcome $m$ to Bob.  Then Bob performs $(u_\B u^m v_\B)^\dag$ depending on Alice's measurement outcome $m$.  Bob's final state becomes $\ket{\psi_\B}$, therefore, LOCC one-piece relocalization has been achieved.  Thus, we have proven that a global unitary operation $U$ is LOCC one-piece relocalizable if and only if $U$ is local unitary equivalent to a controlled-unitary operation.   

\section{Conclusion and discussions}

In this paper, we investigated how originally localized two pieces of quantum information represented by the tensor product of two unknown qudit states is delocalized by performing two-qudit global unitary operations.  To characterize the delocalization power of global unitary operations on quantum information, we analyzed the necessary and sufficient condition to deterministically relocalize one of the two pieces of quantum information to its original Hilbert space by using only LOCC.   We proved that LOCC one-piece relocalization is possible if and only if the global unitary operation is local unitary equivalent to a controlled-unitary operations given by (\ref{localunitaryform}), by reducing general LOCC protocols for one-piece relocalization into one-way LOCC protocols and also a general measurement used in the one-way LOCC protocols into a projective measurement.

For any global unitary operation that is not local unitary equivalent to controlled unitary operations, our result implies that additional non-local resource (entanglement) on top of LOCC is required to perform the one-piece relocalization.   We can interpret this result that such global unitary operations have more delocalization power on quantum information than controlled-unitary operations, therefore, they strongly ``intertwine'' the two pieces of quantum information such that LOCC cannot untwine even sacrificing a piece of quantum information.  

Our result also reveals that the delocalization power and the entangling power \cite{KrausCirac} characterize different non-local properties of global unitary operations. To see this, we consider a two-qubit global unitary operation given by
\begin{equation}
\exp (\rmi \alpha \sum_{j=x,y,z} \sigma_{\A}^j \otimes \sigma_{\B}^j),
\label{class2state}
\end{equation}
which is in the form of decomposition proposed by \cite{KrausCirac}, 
where $\sigma_{\A}^j$ and $\sigma_{\B}^j$ ($j=x,y,z$) are Pauli operators on Alice's Hilbert space and Bob's Hilbert space, respectively.  The operator Schmidt rank \cite{OSR} of this global unitary operation is known to be four unless $\alpha=0$, and that of a controlled-unitary operation is known to be two unless it is the identity operation.  Two-qubit unitary operations are local unitary equivalent to each other if and only if their operator Schmidt ranks coincide.  Thus, the global unitary operation given by (\ref{class2state}) for an arbitrary nonzero $\alpha$  is {\it not} LOCC one-piece relocalizable, and belongs to the higher class in terms of the delocalization power.  The entangling power depends on the parameter $\alpha$ and, as $\alpha$ tends to zero, it reaches to zero.   So for very small $\alpha$, the entangling power of (\ref{class2state}) can be arbitrarily close to zero.   On the other hand, a CNOT (controlled-$\sigma_\B^x$) operation given by $\ket{0} \bra{0} \otimes \mathbb{I} + \ket{1} \bra{1} \otimes \sigma^x_B$ is LOCC one-piece relocalizable, therefore it belongs to the lower class in terms of the delocalization power, but it has higher entangling power (1-ebit) than (\ref{class2state}) for very small $\alpha$. 

It is also possible to further quantify the delocalization power of global unitary operations that are not local unitary equivalent to controlled-unitary operations, by evaluating how much additional ``non-localness'' is required to perform the one-piece relocalization map.   One way to quantify such non-localness is to use the ``entangling power''  $f_{\mathrm{ep}}(\Gamma)$ of a map $\Gamma$, \ie
\begin{equation}
 f_{\mathrm{ep}}(\Gamma) = \max_{\rho_{\mathrm{AB}} \in \mathcal{S}(\mathscr{H}_\A \otimes \mathscr{H}_\B)} E(\Gamma(\rho_{\mathrm{AB}}))-E(\rho_{\mathrm{AB}}),
\end{equation}
where $E(*)$ is some entanglement measure.  Another possible way is to consider the implementation of one-piece relocalization via entanglement-assisted LOCC and evaluate the minimum amount of assisting entanglement.  More formally, this ``entanglement cost'' $f_{\mathrm{c}}(\Gamma)$ for entanglement-assisted LOCC implementation of a map $\Gamma$ is defined by
\begin{equation}
 f_{\mathrm{c}}(\Gamma) =  \min \{E(\rho_{\mathrm{res}})|{}^\exists \Lambda_{\mathrm{LOCC}}
 s.t.~\Tr_{\mathrm{res}} \Lambda_{\mathrm{LOCC}}(\rho_{\mathrm{AB}}\otimes\rho_{\mathrm{res}})=\Gamma(\rho_{\mathrm{AB}})\},
\end{equation}
where $\rho_{\mathrm{res}}$ denotes the assisting entanglement resource shared between the two parties for implementing the map $\Gamma$.  We leave this direction of analysis for future works. 

Finally, we stress that our result is on the delocalization power of global unitary operations for {\it two pieces} of input quantum information.  As we have described in Section 1, if Alice's qudit state is a known pure state and we investigate the delocalization power of a global unitary operation only on Bob's {\it piece} of input quantum information, this problem corresponds to the work of Gregoratti and Werner \cite{GW}.  They have shown that the reduced map (channel) for the one piece of quantum information has to be a random unitary channel.    The Stinespring extension of this random unitary channel can be described by using a controlled-unitary operation.  However, the global unitary operation used in the Stinespring extension of this channel is {\it not necessary} to be a controlled-unitary operation.   For example, a two qubit unitary operation given by $U=\ket{0} \bra{0} \otimes \ket{0} \bra{0} + \ket{0} \bra{1} \otimes \ket{1} \bra{0} + \ket{1} \bra{0} \otimes \ket{0} \bra{1} - \ket{1} \bra{1} \otimes \ket{1} \bra{1}$ has been shown that it is {\it not} local unitary equivalent to a controlled-unitary operation in \cite{ADQC}.   But if we fix Alice's state to be  $\ket{+}= \left ( \ket{0} + \ket{1} \right ) / \sqrt{2}$, LOCC one-piece relocalization can be achieved by Alice's projective measurement $\left \{ \ket{+} \bra{+}, \ket{-} \bra{-} \right \}$ followed by Bob's local unitary operation $H$ when Alice's outcome is $+$, or $\sigma_{\B}^z \cdot H$ when it is $-$, where $H$ denotes the Hadamard gate given by $H=\ket{0} \bra{+} + \ket{1} \bra{-}$.   Therefore,  the delocalization power of global unitary operations depends on the number of pieces of input quantum information.    

\section*{Acknowledgements}  The authors thank P.S. Turner for helpful discussions.   This work is supported by Special Coordination Funds for Promoting Science and
Technology.


\end{document}